\documentclass[10pt]{article}
\usepackage{amsmath}
\usepackage{amssymb}
\usepackage{graphicx}
\usepackage{cite}
\usepackage{color} 
\usepackage[labelfont=bf,labelsep=period,justification=raggedright]{caption}

\makeatletter
\renewcommand{\@biblabel}[1]{\quad#1.}
\makeatother

\date{}

\begin{document}

\begin{flushleft}
{\Large
\textbf{If Cooperation is Likely Punish Mildly: Insights from Economic Experiments Based on the Snowdrift Game}
}\sffamily
\\[3mm]
\textbf{Luo-Luo Jiang,$^{1,2,\ast}$ Matja{\v z} Perc,$^{3}$ Attila Szolnoki$^{4}$}
\\[2mm]
{\bf 1} College of Physics and Electronic Information Engineering, Wenzhou University, Wenzhou, China
{\bf 2} Financial Research Institute, Wenzhou University, Wenzhou, China {\bf 3} Department of Physics, Faculty of Natural Sciences and Mathematics, University of Maribor, Maribor, Slovenia
{\bf 4} Institute of Technical Physics and Materials Science, Research Centre for Natural Sciences, Hungarian Academy of Sciences, Budapest, Hungary
\\[2mm]
$^{\ast}$jiangluoluo@gmail.com
\end{flushleft}
\sffamily

\section*{Abstract}
Punishment may deter antisocial behavior. Yet to punish is costly, and the costs often do not offset the gains that are due to elevated levels of cooperation. However, the effectiveness of punishment depends not only on how costly it is, but also on the circumstances defining the social dilemma. Using the snowdrift game as the basis, we have conducted a series of economic experiments to determine whether severe punishment is more effective than mild punishment. We have observed that severe punishment is not necessarily more effective, even if the cost of punishment is identical in both cases. The benefits of severe punishment become evident only under extremely adverse conditions, when to cooperate is highly improbable in the absence of sanctions. If cooperation is likely, mild punishment is not less effective and leads to higher average payoffs, and is thus the much preferred alternative. Presented results suggest that the positive effects of punishment stem not only from imposed fines, but may also have a psychological background. Small fines can do wonders in motivating us to chose cooperation over defection, but without the paralyzing effect that may be brought about by large fines. The later should be utilized only when absolutely necessary.

\section*{Introduction}
Approximately two million years ago some hominids were beginning to evolve larger brains and body size and to mature more slowly than other apes \cite{calder_84}. This likely procreated serious challenges in rearing offspring that survived. Faced with such evolutionary pressures, members of the genus \textit{Homo} begun helping each other, in particularly provisioning for the young of others regardless of kinship \cite{hrdy_11}. Today, we are known as the supercooperators \cite{nowak_11}, and it is beyond doubt that selfless cooperative behavior between unrelated individuals is one of the key pillars of our remarkable evolutionary success. The temptations to defect, however, have been present in the past as they are now, and we are well aware of the fact that defection may lead to the tragedy of the commons \cite{hardin_g_s68}. But since we are no longer threatened by other species -- in fact, it seems difficult to dispute that the biggest challenges today are of our own production  -- the primal motive to cooperate is gone. We must rely on our cultural heritage and upbringing as well as between-group conflicts to maintain in-group solidarity \cite{bowles_11}.

Perhaps not surprisingly, we have come to appreciate actions that may promote cooperation, most notably punishment \cite{sigmund_tee07, sigmund_10}, even to the point of institutionalization \cite{gurerk_s06, henrich_s06, sigmund_n10, szolnoki_pre11, traulsen_prsb12, isakov_dga12}. The problem is that punishment is costly, and it is far from clear who should be the ones to pay. We can be quick to conclude that obviously it would be on the cooperators to trace down and punish defectors. Yet cooperators already have a personal disadvantage over defectors, and adding yet another could prove too much to bare in a competitive environment. The emergence of second-order free-riding, i.e., contributing to the common pool but not to sanctioning, therefore seems inevitable \cite{boyd_pnas03, helbing_ploscb10, amor_pre11, perc_srep12, deng_k_tpb12, hilbe_srep12}, and in fact presents the biggest threat to the success of punishment \cite{panchanathan_n04, fehr_n04, fowler_n05b}. The hope, or rather the assumption, is that in the long run punishment would pay off, so that the additional investments would be offset by increased levels of cooperation. There exists evidence, both theoretical and experimental, in support of such an assumption \cite{fehr_aer00, fehr_n02, egas_prsb08, gaechter_s08, boyd_s10, perc_njp12, espin_prsb12}, but there are also studies asserting that costly punishment is maladaptive \cite{dreber_n08, rand_jtb09}, and that it can be challenged by antisocial punishment \cite{herrmann_s08, rand_jtb10, gachter_eer11, rand_nc11} as well as reward \cite{rand_s09}. The stick versus carrot dilemma \cite{andreoni_aer03} has recently received ample attention \cite{sefton_ei07, szolnoki_epl10, hilbe_prsb10, hauert_jtb10, sutter_res10, szolnoki_njp12, choi_jep13, vukov_pcbi13}, and the subject of antisocial punishment has also been contested with loners \cite{garcia_jtb12}, which were originally studied in \cite{hauert_s02, hauert_s07}. It is safe to conclude that there are still many open issues that require further research.

Here we investigate the impact of punishment from a somewhat different perspective, namely how humans react when being subject to punishment. When a defector is punished, she essentially has two options on how to proceed. One is to keep defecting in the hope that the withdrawn contribution to the common pool will make up for future sanctions, while second is to decide to cooperate and thus avoid further sanctions. It is a dilemma that is likely to be decided based on the severity of punishment as well as the cost-to-benefit ratio of the game. Punishment can be considered as being effective if a high fraction of punished defectors chooses to cooperate in the next round. To clarify this, we have conducted economic experiments \cite{camerer_03} (recent examples of research are \cite{grujic_pone10, grujic_pone12, grujic_srep12, gracia-lazaro_pnas12, gracia-lazaro_srep12, rand_n12}) based on the snowdrift game played in groups \cite{santos_md_jtb12}, where the two main free parameters were the severity of punishment and the cost-to-benefit ratio. In the realm of the game (see Methods for details), the cost-to-benefit ratio determines just how severe the social dilemma is. Low cost-to-benefit ratios constitute lenient conditions for the evolution of cooperation, while high costs and low benefits favor defection. As we will show, the effectiveness of different punishment regimes depends sensitively on the cost-to-benefit ratio. If costs are low, the application of severe punishment is not more effective than mild punishment, yet it does lead to lower overall payoffs and hence is not recommended. Only if costs are high does severe punishment outperform mild punishment in terms of persuading more defectors to adopt cooperation in the next round. We proceed by presenting the main results in support of these conclusions, first by focusing on the outcome of the game in the absence and subsequently in the presence of punishment.

\section*{Results}
The impact of punishment on the outcome of the game can be understood well only if the same economic experiments are carried out also without the possibility of this action. We therefore first conduct experiments in the absence of punishment to arrive at a baseline scenario, in particular to estimate the general willingness of players to cooperate at different values of the cost-to-benefit ratio. Subsequently, we will use this as a reference point for the snowdrift game with punishment.

\subsection*{The snowdrift game without punishment}
To illustrate the snowdrift game, imagine two drivers that are caught in a blizzard and trapped on either side of a snowdrift~\cite{maynard_82}. They can either get out and start shoveling (cooperate) or remain in the car (defect). If both cooperate, they have the benefit $b$ of getting home while sharing the labor $c$. Thus, $R=b-c/2$, which indicates the $Reward$ for both cooperators. If both defect, they do not get anywhere and hence incur the punishment $P=0$. If only one shovels, however, they both get home but the defector avoids the labor cost and gets the $Temptation$ $T=b$, whereas the cooperator gets the $Sucker's$ payoff $S=b-c$. The four payoff values in the snowdrift game rank in order: $T > R > S >P$, and $r=c/(2b-c)$ illustrates the cost-to-benefit ratio. If we fix $R=b-c/2=1$, then $r=c/(2b-c)=c/2$, $T=1+r$ and $S=1-r$, and the payoff matrix thus becomes:
\begin{equation}
\begin{array}{ccc}
\ & \begin{array}{cc} \textbf{C} &~~~~\textbf{D} \end{array} \\
\begin{array}{c} \textbf{C} \\ ~\textbf{D} \end{array}&\left(\begin{array}{cc} 1&1-r\\
1+r&0
\end{array}\right)
\end{array}
\label{eq:payoff}
\end{equation}

\begin{figure}
\begin{center}\includegraphics[width=11cm]{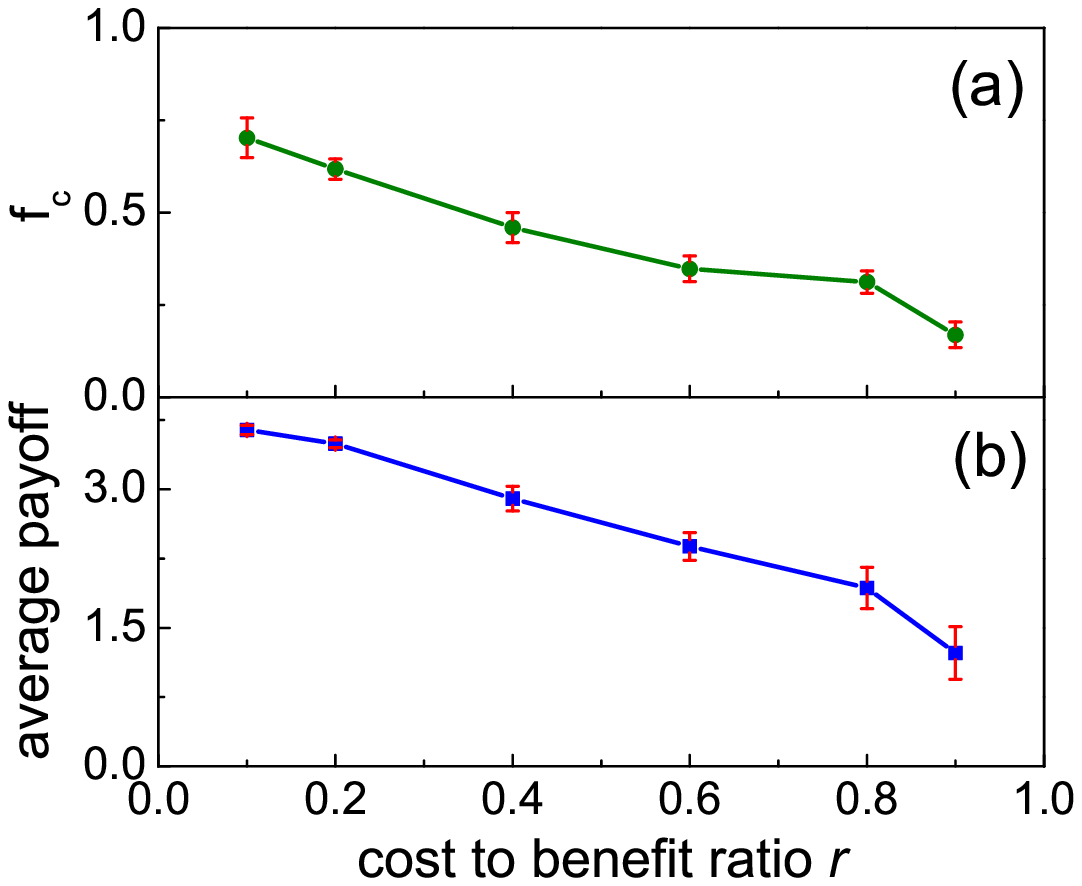}\end{center}
\caption{{\bf Higher cost-to-benefit ratios in the snowdrift game lead to lower levels of cooperation}. Depicted are results of an economic experiment, as obtained in the absence of punishment. Panels (a) and (b) show the frequency of cooperation $f_c$ and the average payoff per period in dependence on the cost-to-benefit ratio $r$, respectively. The whiskers in panels (a) and (b) show the 95\% confidence intervals for the frequency of cooperation and for the average payoffs, respectively.
 \label{fig:nopunishment}}
\end{figure}

To have a reference point for the actual impact of punishment, as noted, we first study how the frequency of cooperation varies with the cost-to-benefit ratio $r$. In the $Treatment~I$ including eight sessions, cooperators were not allowed to punish defectors, and six different values of cost-to-benefit ratio $r$ were set (four sessions for $r=0.2$ and $r=0.8$, other four sessions for other values of $r$). In each session 20 undergraduate students were randomly allocated to four groups of five subjects playing snowdrift games in the Computer Lab for Behavior Games [for further information see the Methods and Fig.~S1 (Supplementary Information)]. As it is expected, both the level of cooperation $f_{c}$ and the average payoff per period decrease with increasing of $r$, as demonstrated in Figs.~\ref{fig:nopunishment}~(a) and (b), respectively. These results signal clearly the conflict between the individual and group interests: the best strategy for individuals is to defect if the opponent adopts cooperation. Consequently, the total payoff of the whole group falls gradually by increasing $r$, which prompts more individuals to choose defection to gain a higher individual payoff. In the end, when the cost-to-benefit ratio $r$ is large, both individual and the group benefits become minimal. The Kruskal-Wallis test for results presented in Fig.~\ref{fig:nopunishment}~(a) yields $\chi_{5}^{2}=53.923$ and $P=0.0001$ for individual-level data, and $\chi_{5}^{2}=27.782$ and $p=0.0001$ for group-level data. For Fig.~\ref{fig:nopunishment}~(b), on the other hand, we obtain $\chi_{5}^{2}=131.103$ and $p=0.0001$ for individual-level data, and $\chi_{5}^{2}=27.757$ and $P=0.0001$ for group-level data. In addition, we present details of the regression analysis for results presented in Fig.~\ref{fig:nopunishment} in Tables S1, S2, S3 and S4 (Supplementary Information). This statistical analysis indicates clearly that the cost-to-benefit ratio indeed does have a statistically significant impact on the cooperation level and the average payoff.

\begin{figure}
\begin{center}\includegraphics[width=11cm]{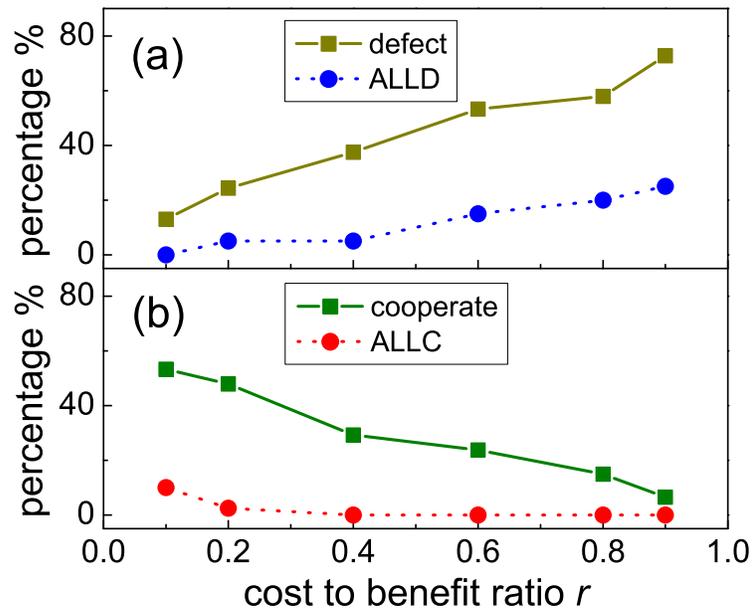}\end{center}
\caption{{\bf Statistics on when cooperators continue to cooperate and defectors continue to defect, and vice versa, in the absence of punishment}. Panel (a) shows the percentage of defectors choosing defection in the next round and the percentage of individuals who always defect in dependence on the cost-to-benefit ratio $r$. Panel (b) shows the percentage of cooperators choosing cooperation in the next round and the percentage of individuals who always cooperate in dependence on the cost-to-benefit ratio $r$. \label{fig:nopunishmentcontinu}}
\end{figure}

In addition to the average values of strategies we also monitored how often players change strategies at different values of $r$. We found that defectors adhere to defection as we increase $r$. Cooperators, on the other hand, do not insist on cooperation at high $r$. Figures~\ref{fig:nopunishmentcontinu}~(a) and (b) show the percentage of defectors selecting defection and cooperators selecting cooperation in the next round among all the subjects that changed strategies in the next round. We found that the percentage of defectors choosing to stick with defection increases from 13\% at $r=0.1$ to 72.75\% at $r=0.9$, while that of cooperators choosing to cooperate anew declines from 53.25\% at $r=0.1$ to 6.5\% at $r=0.9$. Furthermore, we have also monitored the fraction of individuals who {\it always} choose to cooperate (ALLC) and those who always choose to defect (ALLD) among all the subjects. Interestingly, these values depend sensitively on the cost-to-benefit ratio. As the dotted lines in Fig.~\ref{fig:nopunishmentcontinu} show, the percentage of ALLD increases greatly from 0\% at $r=0.1$ to 25\% at $r=0.9$, while the percentage of ALLC reduces from 10\% at $r=0.1$ to 0\% suddenly for higher $r$ values. These observations suggest that the subjects are in general ``flexible'' in responding to the change of external conditions (here determined by the value of $r$), and are thus well aware and concerned for their individual success.

\subsection*{The snowdrift game with punishment}
To investigate the effectiveness of punishment when applying different fines, we focus on $r=0.2$ and $r=0.8$, because these two values of the cost-to-benefit ratio represent typical conditions constituting low and high costs of cooperation, respectively (see Figs.~\ref{fig:nopunishment} and \ref{fig:nopunishmentcontinu}). For the sake of simplicity, there were two stages making up sessions $Treatment~II$ and $Treatment~III$, namely the playing game stage and the peer punishment stage. During the first stage, subjects played the snowdrift game with other group members. Similarly as in $Treatment~I$, there were four groups containing five subjects each. During the second stage, cooperators were given the chance to punish defectors on a peer-to-peer (individual) basis as follows. If there was at least one cooperator who accepted the cost of punishment, then all the payoffs of all defector in the group were reduced by a fine $p$, and simultaneously the punisher's profit was also reduced by a single unit, which was the costs of punishment. We should stress that the cost of punishment was always constant at different values of fine. Therefore the results we observed primarily focus on the reaction of defectors being punished and not the dilemma of cooperators whether to punish or not. The latter dilemma, however, still exist because those who cooperate but do not punish can be considered as second free-riders. Table~\ref{tal:parameters} illustrates the applied parameter values for $Treatment~II$ and $Treatment~III$.

\begin{table}[!ht]
\caption{\footnotesize {\bf Game parameters employed during $\textit{Treatment~II}$ and $\textit{Treatment~III}$ in the snowdrift game with punishment}. $Treatment~II$ corresponds to mild punishment because the fines ($p$) for defection are low, while $Treatment~III$ corresponds to severe punishment because of the application of comparatively high penalties.} \label{tal:parameters}
\begin{tabular}{ccccc}
\hline
$SD~game~conditions$~~&~~~~~~$low~~cost$~~~~~~& $high~~cost$\\
\hline
$Treatment~II$~~~& $r=0.2$,~$p=2.0$;~~& $r=0.8$,~$p=2.0$\\
\hline
$Treatment~III$~~& $r=0.2$,~$p=4.0$;~~~& $r=0.8$,~$p=4.0$ \\
\hline
\end{tabular}
\end{table}

\begin{figure}
\begin{center}\includegraphics[width=16cm]{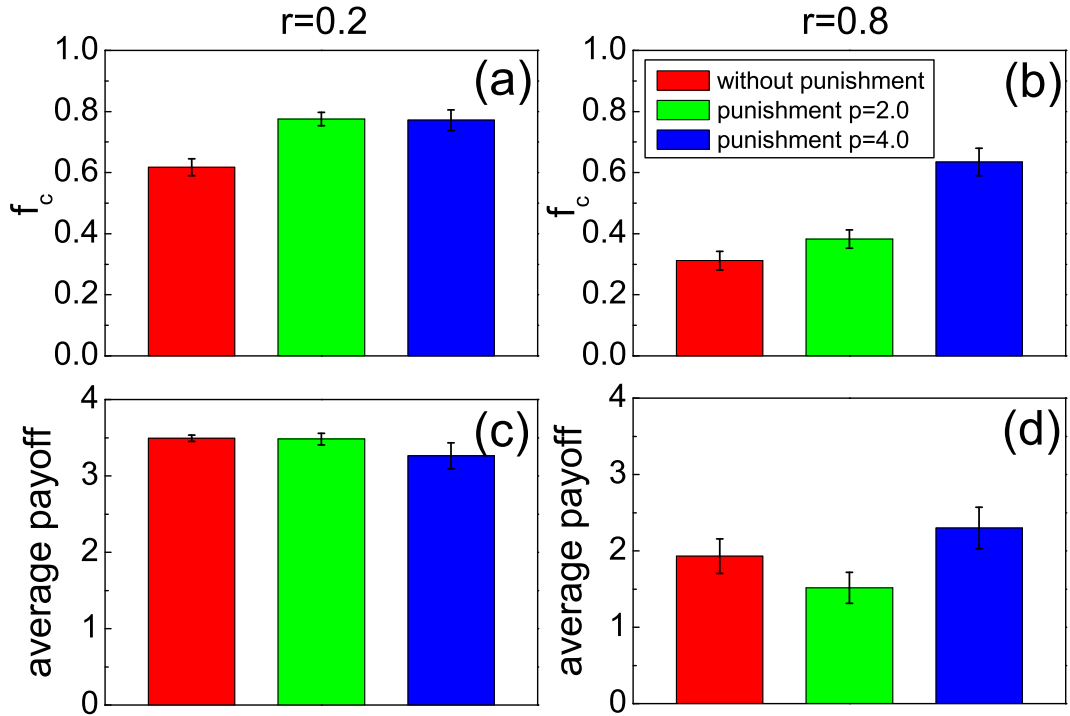}\end{center}
\caption{{\bf Mild punishment outperforms severe punishment if the conditions for cooperation are favorable}. Panels (a) and (c) show the frequency of cooperation and the average payoff per period in the absence of punishment, for punishment with $p=2.0$, and for punishment with $p=4.0$, as obtained when the cost-to-benefit ratio is $r=0.2$ (low). Panels (b) and (d) show the frequency of cooperation and the average payoff per period in the absence of punishment, for punishment with $p=2.0$, and for punishment with $p=4.0$, as obtained when the cost-to-benefit ratio is $r=0.8$ (high). Only if cooperation is very unlikely in the absence of sanctions does severe punishment reveal its advantages. The whiskers in panels (a) and (b) show the 95\% confidence intervals for the frequency of cooperation, while the whiskers in panels (c) and (d) show the same for the average payoffs.
\label{fig:punishment}}
\end{figure}

When the cost-to-benefit ratio is low ($r=0.2$), the application of larger fines does not yield a more favorable outcome than the application of lower fines. Naturally, the chance to punish defectors will improve the cooperation level but, as Fig.~\ref{fig:punishment}~(a) shows, higher fines will not increase $f_C$ further. On the other hand, the average payoff of group members will be reduced, especially so for severe punishment, as illustrated in Fig.~\ref{fig:punishment}~(c). Beside the Kruskal-Wallis test we have also calculated the 95\% confidence intervals to compare the impacts of punishment at low $r$. While $f_C$ is in the (0.708,0.846) interval for $p=2.0$, and in (0.653,0.889) for $p=4.0$, signaling not detectable differences, the average payoffs are definitely different: the corresponding confidence interval is (3.407,3.557) for $p=2.0$, and (3.093,3.404) for $p=4.0$. In agreement with previous observations \cite{dreber_n08}, the uselessness of too hard punishments could be an important message for those who are in a position to establish the means of punishment in our society.

Interestingly, the relevance of severe punishment becomes more prominent when the external conditions to cooperate become significantly worse. When the cost-to-benefit ratio is high ($r=0.8$), the application of higher fines will gradually elevate the cooperation level beyond what could be achieved by mild punishment. As shown in Fig.~\ref{fig:punishment}~(b), $f_C$ can be doubled due to severe punishment. The impact on the average payoff is also positive, given that cooperation is virtually absent in the absence of punishment. This is hence a very much desired outcome one would expect from punishment. It is important to emphasize, however, that low fines will still yield a similar outcome as we have observed for $r=0.2$. Namely, as Fig.~\ref{fig:punishment}~(d) clearly illustrates, the usage of $p=2.0$ increases $f_C$, but the average payoff is lower than in the punishment-free case. The corresponding 95\% confidence interval is (1.717,2.154) for the punishment-free case and (1.311,1.707) for $p=2.0$.

Results presented thus far indicate that the value of fine should be carefully adjusted in agreement with the general conditions that characterize the severity of the social dilemma the players are facing. If the conditions are such that cooperation is likely and viable even in the absence of punishment, then severe sanctioning of defective behavior should be avoided as it leads to lower average payoffs. On the other hand, in strongly defection-prone environments, where cooperators hardly have a chance to survive in the absence of additional regulations, severe punishment appears to be the correct and indeed the only effective means to evoke a change for the better.

\begin{figure}
\begin{center}\includegraphics[width=13cm]{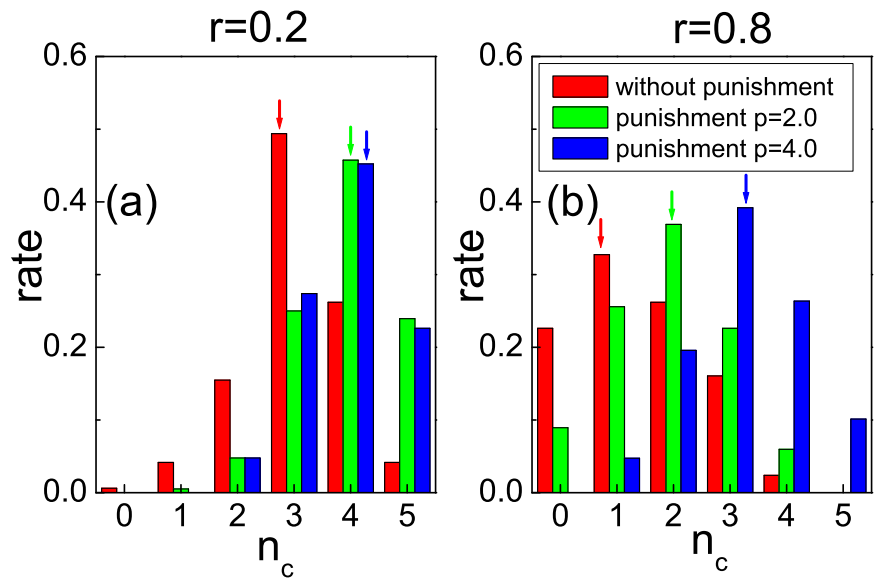}\end{center}
\caption{{\bf Distribution of strategies within groups depends not just on the severity of punishment, but also on the severity of the social dilemma}. Panels (a) and (b) depict the rate of groups having $n_c$ cooperators, as obtained for $Treatment~I$ (without punishment), $Treatment~II$ (punishment with $p=2.0$), and $Treatment~III$ (punishment with $p=4.0$), at $r=0.2$ and $r=0.8$, respectively. Only if $r=0.8$ is severe punishment more effective. If cooperation is likely ($r=0.2$) mild punishment is at least just as effective in sustaining highly cooperative groups. \label{fig:rate}}
\end{figure}

To reveal the microscopic details governing the choices during the conducted economic experiments, we have also determined the rate of groups that had different numbers of cooperators during all the considered periods. For example, there were $83$ groups with three cooperators in the total of $168$ groups for all the periods in the absence of punishment at $r=0.2$. Therefore the rate of groups with three cooperators is $83/168\approx0.49$. This is the most common formation, as shown in Fig.~\ref{fig:rate}~(a). If punishment was applied at $r=0.2$, then the most typical group would contain four cooperators and one defector to form the five group members. Fig.~\ref{fig:rate}~(a) also illustrates that not just the average $f_C$ but also the probability distributions of different groups are very similar when we applied different punishment strengths. This explains why a more severe punishment is less recommended in this case: it has no additional impact on the strategy choice of players, and hence it only contributes to additionally reducing the payoffs of defectors. This may have a negative psychological side effect, as it makes it less likely that such ``paralyzed'' players will attempt reintegration by means of the less profitable cooperative approach.

If cooperation is costly, however, the distribution of strategies within the groups changes significantly, as shown in Fig.~\ref{fig:rate}~(b). Here, the most common group contains only a single cooperator in the absence of punishment. Moreover, there is a significant fraction of groups, about 20\% of them, which completely fulfill the makings of the tragedy of the commons as therein everybody chooses to defect. If we apply punishment then the number of cooperators $n_c$ increases, and indeed the maximum of the distribution moves towards higher $n_c$. In particular, it is at $n_c=2$ when $p=2.0$ and at $n_c=3$ for $p=4.0$. In agreement with our previous conclusion, here the application of severe punishment will significantly reduce the number of defectors, and this reduction comfortably makes up for the losses in overall payoff that are charged to defectors because of the larger value of $p$.

\begin{figure}
\begin{center}\includegraphics[width=14cm]{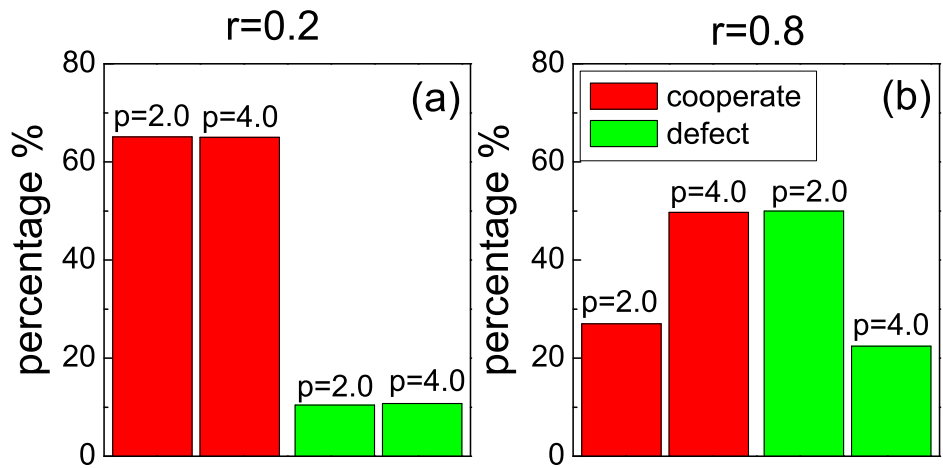}\end{center}
\caption{{\bf If to cooperate is a very difficult proposition, severe punishment is more likely to divert from defection and perpetuate cooperation than mild punishment}. Depicted is the statistics on when cooperators continue to cooperate (red) and defectors continue to defect (green) under mild ($p=2.0$) and severe ($p=4.0$) punishment. In panel (a), for $r=0.2$, mild punishment is just as effective as severe punishment in maintaining the strategy choices. In panel (b), for $r=0.8$, mild punishment is less effective. If punishment is severe, cooperators are more likely to continue cooperating (red), while defectors are less likely to continue defecting (green) than if punishment is mild. \label{fig:punishmentcontinu}}
\end{figure}

To arrive at addressing our principal goal, which regards the effectiveness of punishment, we have also determined the percentage of defectors who select defection and cooperators who choose cooperation in the next round among all players who change strategy. Figures~\ref{fig:punishmentcontinu}~(a) and (b) show these ratios for low and high costs of cooperation, i.e., $r=0.2$ and $0.8$, respectively. In comparison with the data obtained without punishment (plotted in Fig.~\ref{fig:nopunishmentcontinu}), we found that for $r=0.2$ there is a slight increase in opting to cooperate and decrease in opting to defect, but the value of fine plays a rather insignificant role in mediating this decision. As we have already observed, this changes significantly if a high cost-to-benefit ratio characterizes the snowdrift game. Here the percentage of cooperators staying cooperators increases from 27.0\% for $p=2.0$ to 49.7\% for $p=4.0$, and the percentage of defectors deciding to defect again reduces from 50.4\% for $p=2.0$ to 22.4\% for $p=4.0$. For a deeper insight we have also calculated how the probability of changing strategy depends on the actions of the others in the group (on the number of cooperative opponents) at different values of $r$ values, as summarized in Fig.~S2 (Supplementary Information). This further strengthens the conclusion that the proper impact of punishment on individual decision making might depend sensitively on other elementary circumstances, like in our case, how beneficial it is to defect instead of to cooperate to begin with.

\begin{figure}
\begin{center}\includegraphics[width=14cm]{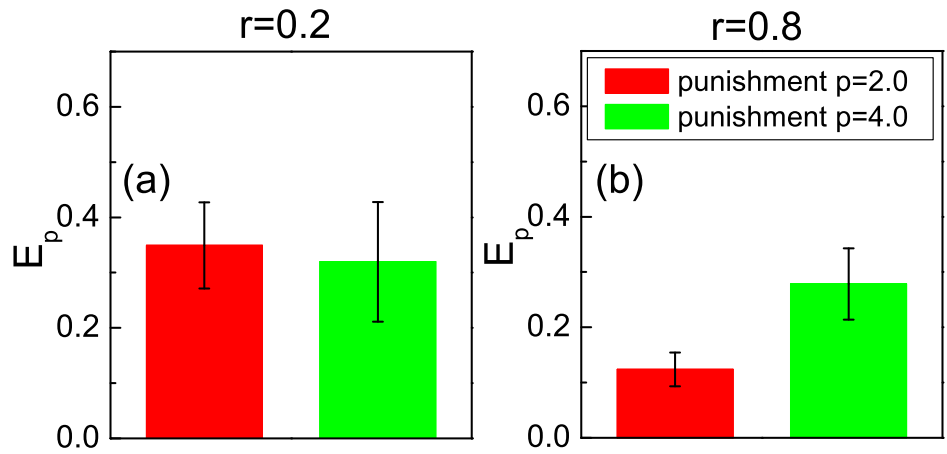}\end{center}
\caption{{\bf If cooperation is likely the effectiveness of mild punishment is just as high as the effectiveness of severe punishment}. Panels (a) and (b) present results obtained for $r=0.2$ and $r=0.8$, respectively. It can be observed that for $r=0.2$, when the likelihood of cooperation is high even in the absence of sanctioning (see Fig.~\ref{fig:nopunishment}), mild punishment is just as effective as severe punishment. Conversely, for $r=0.8$ severe punishment leads to a higher percentage of defectors that after being punished choose to cooperate ($E_p$) than mild punishment. The whiskers in panels (a) and (b) show the 95\% confidence intervals for the effectiveness of punishment $E_p$.
\label{fig:effective}}
\end{figure}

Staying further at the high cost regime, we note that it is difficult to distinguish accurately the motivation of a defector to choose cooperation in the next round, because the fluctuations of a person to choose a different strategy in the next round amount to about 12\%, as shown in Fig.~S3 (Supplementary Information). We argue that the primary purpose of punishment ought to be to turn defectors into cooperators at the next time of asking. As shown in Fig.~S4 (Supplementary Information), the percentage of defectors who choose cooperation in the next round because of being punished in the current round increases from 8\% to 11\% when the fine is increased. To test this further, we can define the effectiveness of punishment $E_{p}$ as the rate of defectors who choose cooperation in the next round after being punished. As shown in Fig.~\ref{fig:effective}~(a), the effectiveness of punishment using $p=2.0$ is equal or even a bit larger than that of $p=4.0$ when the cost-to-benefit ratio is low ($r=0.2$). Since here two rather than three setups are tested against statistical relevance, we apply the t-test, which yields $P>0.05$. This indicates that, for $r=0.2$, there are indeed no statistically relevant differences between the effectiveness of punishment with $p=2.0$ and $p=4.0$. The difference at $r=0.8$, depicted in Fig.~\ref{fig:effective}~(b), is rather more spectacular. There the higher fine is much more effective in converting defectors to cooperators, and indeed it corroborates the necessity of severe punishment in adverse environments. Here the t-test yields $P<0.0001$, clearly confirming statistically relevant differences between the two punishment modes at $r=0.8$.

\section*{Discussion}
We have conducted economic experiments centered around the snowdrift game played in groups of five, with the aim of determining the effectiveness of severe and mild punishment to persuade defectors to choose cooperation in the next round of the game. With the assumption that the propensity of the environment itself to promote or deter cooperation likely plays an important role, we have tested the impact of severe and mild punishment under a cooperation-prone and under a defection-prone setup of the snowdrift game. We have observed that benefits of severe punishment emerge only under adverse conditions, when to cooperate is highly unlikely in the absence of sanctions. If the conditions are favorable or at least not unfavorable, mild punishment is not less effective. In particular, if cooperation is likely, mild punishment is just as effective as severe punishment in persuading defectors to choose cooperation. But since the fines imposed by severe punishment are higher, the overall welfare is lower than by mild punishment. Severe punishment fails to offset the imposed fines and costs associated with its execution, and so the players would be better of without it. Importantly, this holds even under the lenient assumption that the cost of punishment is independent of the severity of punishment. If the costs would scale with the imposed fines, the effectiveness of severe punishment would be even worse. However, if the conditions for cooperation are unfavorable, then only severe punishment is able to revert the players from defecting, and it is also then that it has a positive impact on the average payoff and is in fact sustainable.

The presented results indicate that it is far from obvious to know how large fines should be applied to elevate the overall welfare, even if the costs do not scale with the imposed penalties. Contrary to what could be assumed, even if we have the means to punish hard, doing so is likely not an optimal decision. It can be a viable one if the conditions are really adverse and unfavorable for the evolution of cooperation. In general, however, mild punishment is not less effective as severe punishment, with the added benefit that the imposed fines make it easier for the punished individuals to reintegrate into the society. In view of these observations, we conclude that the positive effects of punishment stem not only from the imposed fines, but may also have a psychological background. Small fines work just as well as high fines in motivating us to chose cooperation over defection. Punishing excessively hard seldom has additional benefits, but it does have the potential to disable the punished individual, and it also decreases the overall welfare more than punishing mild or moderately. Neither of these two side effects is desirable, and thus we conclude that severe punishment should be utilized only when absolutely necessary. It seems less harm can be done by adopting mild punishment and risking a few more persistent defectors, then it is to endorse severe punishment in the name of total cooperation.

\section*{Methods}
A total of $320$ undergraduate students (45\% females, $20.3$ years old on average) from Wenzhou University participated in repeated snowdrift games taking place in groups of five at the Computer Lab of Behavior Games. Students that participated did so by answering a public call that was issued by the Computer Lab of Behavior Games of Wenzhou University. The ethics committee of the Wenzhou University approved the public call and the experiments. All the participants provided their written informed consent to participate in the study. Prior to participation, they have also learned the rules of the game and subsequently demonstrated their understanding in a short test.

The 20 subjects in a session were allocated anonymously to four groups consisting of five subjects each by means of the z-Tree software~\cite{fischbacher_ee07}. Subsequently, subjects played the snowdrift game with all the members in the same group. Since participants were freshmen and sophomore students with different major fields, coming from different departments, they were unlikely to know each other. In addition, subjects were not allowed to participate in more than a single session of the experiment. A total of sixteen sessions were conducted from May to December 2012. Three different treatments were conducted. Namely, experiments without punishment for different cost-to-benefit ratios $r$ ($Treatment~I$), experiments with punishment with penalty $p=2.0$ for $r=0.2$ and $r=0.8$ ($Treatment~II$), and experiments with punishment with penalty $p=4.0$ for $r=0.2$ and $r=0.8$ ($Treatment~III$). Each subject played $25$ periods during approximately $60$ min, and earned 64.8 RMB (the Chinese unit of currency) on average, which amounts to approximately 10.3 US dollars.

At the beginning of each experiment, subjects read written instructions that explained the payoff matrix and the rules of the game. To avoid misunderstanding the instructions, subjects were asked to calculate their own payoff for several examples, and they had to arrive at the correct numbers in order to be allowed participation. After the experiment started, participants marked their decisions on a computer screen using the experimental software z-Tree. In every period, subjects were informed of their own decision and their monetary payoff on the computer screen. Cooperators were allowed to punish defectors during the punishment stage of $Treatment~II$ and $Treatment~III$. Each subject's final score was summed over all periods, and subjects earned an income proportional to their final score (1 RMB for each score point).

\section*{Acknowledgments}
L.-L. J. would like to thank Ke-Zhong Jin for helpful discussions. This work was supported by the National Natural Science Foundation of China (Grants 61203145 and 11047012), the Hungarian National Research Fund (Grant K-101490), and the Slovenian Research Agency (Grant J1-4055).

\clearpage

\begin{thebibliography}{10}
\providecommand{\url}[1]{\texttt{#1}}
\providecommand{\urlprefix}{URL }
\expandafter\ifx\csname urlstyle\endcsname\relax
  \providecommand{\doi}[1]{doi:\discretionary{}{}{}#1}\else
  \providecommand{\doi}{doi:\discretionary{}{}{}\begingroup
  \urlstyle{rm}\Url}\fi
\providecommand{\bibAnnoteFile}[1]{%
  \IfFileExists{#1}{\begin{quotation}\noindent\textsc{Key:} #1\\
  \textsc{Annotation:}\ \input{#1}\end{quotation}}{}}
\providecommand{\bibAnnote}[2]{%
  \begin{quotation}\noindent\textsc{Key:} #1\\
  \textsc{Annotation:}\ #2\end{quotation}}
\providecommand{\eprint}[2][]{\url{#2}}

\bibitem{calder_84}
Calder WA (1984) Size, function, and life history.
\newblock Cambridge, MA: Cambridge University Press.
\input{calder_84}

\bibitem{hrdy_11}
Hrdy SB (2011) Mothers and Others: The Evolutionary Origins of Mutual
  Understanding.
\newblock Cambridge, MA: Harvard University Press.
\input{hrdy_11}

\bibitem{nowak_11}
Nowak MA, Highfield R (2011) SuperCooperators: Altruism, Evolution, and Why We
  Need Each Other to Succeed.
\newblock New York: Free Press.
\input{nowak_11}

\bibitem{hardin_g_s68}
Hardin G (1968) The tragedy of the commons.
\newblock Science 162: 1243--1248.
\input{hardin_g_s68}

\bibitem{bowles_11}
Bowles S, Gintis H (2011) A Cooperative Species: Human Reciprocity and Its
  Evolution.
\newblock Princeton, NJ: Princeton University Press.
\input{bowles_11}

\bibitem{sigmund_tee07}
Sigmund K (2007) Punish or perish? retailation and collaboration among humans.
\newblock Trends Ecol Evol 22: 593--600.
\input{sigmund_tee07}

\bibitem{sigmund_10}
Sigmund K (2010) The Calculus of Selfishness.
\newblock Princeton, NJ: Princeton University Press.
\input{sigmund_10}

\bibitem{gurerk_s06}
Gurerk O, Irlenbusch B, Rockenbach B (2006) The competitive advantage of
  sanctioning institutions.
\newblock Science 312: 108--111.
\input{gurerk_s06}

\bibitem{henrich_s06}
Henrich J (2006) Cooperation, punishment, and the evolution of human
  institutions.
\newblock Science 312: 60--61.
\input{henrich_s06}

\bibitem{sigmund_n10}
Sigmund K, De~Silva H, Traulsen A, Hauert C (2010) Social learning promotes
  institutions for governing the commons.
\newblock Nature 466: 861--863.
\input{sigmund_n10}

\bibitem{szolnoki_pre11}
Szolnoki A, Szab{\'o} G, Perc M (2011) Phase diagrams for the spatial public
  goods game with pool punishment.
\newblock Phys Rev E 83: 036101.
\input{szolnoki_pre11}

\bibitem{traulsen_prsb12}
Traulsen A, R{\"o}hl T, Milinski M (2012) An economic experiment reveals that
  humans prefer pool punishment to maintain the commons.
\newblock Proc R Soc B 279: 3716--3721.
\input{traulsen_prsb12}

\bibitem{isakov_dga12}
Isakov A, Rand D (2012) The evolution of coercive institutional punishment.
\newblock Dyn Games Appl 2: 97--109.
\input{isakov_dga12}

\bibitem{boyd_pnas03}
Boyd R, Gintis H, Bowles S, Richerson PJ (2003) The evolution of altruistic
  punishment.
\newblock Proc Natl Acad Sci USA 100: 3531--3535.
\input{boyd_pnas03}

\bibitem{helbing_ploscb10}
Helbing D, Szolnoki A, Perc M, Szab{\'o} G (2010) Evolutionary establishment of
  moral and double moral standards through spatial interactions.
\newblock PLoS Comput Biol 6: e1000758.
\input{helbing_ploscb10}

\bibitem{amor_pre11}
Amor DR, Fort J (2011) Effects of punishment in a mobile population playing the
  prisoner’s dilemma game.
\newblock Phys Rev E 84: 066115.
\input{amor_pre11}

\bibitem{perc_srep12}
Perc M (2012) Sustainable institutionalized punishment requires elimination of
  second-order free-riders.
\newblock Sci Rep 2: 344.
\input{perc_srep12}

\bibitem{deng_k_tpb12}
Deng K, Li Z, Kurokawa S, Chu T (2012) Rare but severe concerted punishment
  that favors cooperation.
\newblock Theor Popul Biol 81: 284--291.
\input{deng_k_tpb12}

\bibitem{hilbe_srep12}
Hilbe C, Traulsen A (2012) Emergence of responsible sanctions without second
  order free riders, antisocial punishment or spite.
\newblock Sci Rep 2: 458.
\input{hilbe_srep12}

\bibitem{panchanathan_n04}
Panchanathan K, Boyd R (2004) Indirect reciprocity can stabilize cooperation
  without the second-order free rider problem.
\newblock Nature 432: 499--502.
\input{panchanathan_n04}

\bibitem{fehr_n04}
Fehr E (2004) Don't lose your reputation.
\newblock Nature 432: 449--450.
\input{fehr_n04}

\bibitem{fowler_n05b}
Fowler JH (2005) Second-order free-riding problem solved?
\newblock Nature 437: E8--E8.
\input{fowler_n05b}

\bibitem{fehr_aer00}
Fehr E, {G\"a}chter S (2000) Cooperation and punishment in public goods
  experiments.
\newblock Am Econ Rev 90: 980--994.
\input{fehr_aer00}

\bibitem{fehr_n02}
Fehr E, G{\"a}chter S (2002) Altruistic punishment in humans.
\newblock Nature 415: 137--140.
\input{fehr_n02}

\bibitem{egas_prsb08}
Egas M, Riedl A (2008) The economics of altruistic punishment and the
  maintenance of cooperation.
\newblock Proc R Soc B 275: 871--878.
\input{egas_prsb08}

\bibitem{gaechter_s08}
G{\"a}chter S, Renner E, Sefton M (2008) The long-run benefits of punishment.
\newblock Science 322: 1510.
\input{gaechter_s08}

\bibitem{boyd_s10}
Boyd R, Gintis H, Bowles S (2010) Coordinated punishment of defectors sustains
  cooperation and can proliferate when rare.
\newblock Science 328: 617--620.
\input{boyd_s10}

\bibitem{perc_njp12}
Perc M, Szolnoki A (2012) Self-organization of punishment in structured
  populations.
\newblock New J Phys 14: 043013.
\input{perc_njp12}

\bibitem{espin_prsb12}
Esp{\'{\i}}n A, Bra{\~n}as-Garza P, Herrmann B, Gamella J (2012) Patient and
  impatient punishers of free-riders.
\newblock Proc R Soc B .
\input{espin_prsb12}

\bibitem{dreber_n08}
Dreber A, Rand DG, Fudenberg D, Nowak MA (2008) Winners don't punish.
\newblock Nature 452: 348--351.
\input{dreber_n08}

\bibitem{rand_jtb09}
Rand DG, Ohtsuki H, Nowak MA (2009) Direct reciprocity with costly punishment:
  Generous tit-for-tat prevails.
\newblock J Theor Biol 256: 45--57.
\input{rand_jtb09}

\bibitem{herrmann_s08}
Herrmann B, Thoni C, Gachter S (2008) Antisocial punishment across societies.
\newblock Science 319: 1362--1367.
\input{herrmann_s08}

\bibitem{rand_jtb10}
Rand DG, Armao JJ, Nakamaru M, Ohtsuki H (2010) Anti-social punishment can
  prevent the co-evolution of punishment and cooperation.
\newblock J Theor Biol 265: 624--632.
\input{rand_jtb10}

\bibitem{gachter_eer11}
G{\"a}chter S, Herrmann B (2011) The limits of self-governance when cooperators
  get punished: Experimental evidence from urban and rural russia.
\newblock Eur Econ Rev 55: 193--210.
\input{gachter_eer11}

\bibitem{rand_nc11}
Rand DG, Nowak MA (2011) The evolution of antisocial punishment in optional
  public goods games.
\newblock Nat Commun 2: 434.
\input{rand_nc11}

\bibitem{rand_s09}
Rand DG, Dreber A, Ellingsen T, Fudenberg D, Nowak MA (2009) Positive
  interactions promote public cooperation.
\newblock Science 325: 1272--1275.
\input{rand_s09}

\bibitem{andreoni_aer03}
Andreoni J, Harbaugh W, Vesterlund L (2003) The carrot or the stick: Rewards,
  punishments, and cooperation.
\newblock Am Econ Rev 93: 893--902.
\input{andreoni_aer03}

\bibitem{sefton_ei07}
Sefton M, Shupp RS, Walker J (2007) The effects of rewards and sanctions in
  provision of public goods.
\newblock Economic Inquiry 45: 671--690.
\input{sefton_ei07}

\bibitem{szolnoki_epl10}
Szolnoki A, Perc M (2010) Reward and cooperation in the spatial public goods
  game.
\newblock EPL 92: 38003.
\input{szolnoki_epl10}

\bibitem{hilbe_prsb10}
Hilbe C, Sigmund K (2010) Incentives and opportunism: from the carrot to the
  stick.
\newblock Proc R Soc B 277: 2427--2433.
\input{hilbe_prsb10}

\bibitem{hauert_jtb10}
Hauert C (2010) Replicator dynamics of reward \& reputation in public goods
  games.
\newblock J Theor Biol 267: 22--28.
\input{hauert_jtb10}

\bibitem{sutter_res10}
Sutter M, Haigner S, Kocher MG (2010) Choosing the carrot or the stick?
  endogenous institutional choice in social dilemma situations.
\newblock Rev Econ Studies 77: 1540--1566.
\input{sutter_res10}

\bibitem{szolnoki_njp12}
Szolnoki A, Perc M (2012) Evolutionary advantages of adaptive rewarding.
\newblock New J Phys 14: 093016.
\input{szolnoki_njp12}

\bibitem{choi_jep13}
Choi JK, Ahn TK (2013) Strategic reward and altruistic punishment support
  cooperation in a public goods game experiment.
\newblock Journal of Economic Psychology 35: 17--30.
\input{choi_jep13}

\bibitem{vukov_pcbi13}
Vukov J, Pinheiro F, Santos F, Pacheco J (2013) Reward from punishment does not
  emerge at all costs.
\newblock PLoS Comput Biol 9: e1002868.
\input{vukov_pcbi13}

\bibitem{garcia_jtb12}
Garc{\'{\i}}a J, Traulsen A (2012) Leaving the loners alone: Evolution of
  cooperation in the presence of antisocial punishment.
\newblock J Theor Biol 307: 168--173.
\input{garcia_jtb12}

\bibitem{hauert_s02}
Hauert C, De~Monte S, Hofbauer J, Sigmund K (2002) Volunteering as \protect{Red
  Queen} mechanism for cooperation in public goods game.
\newblock Science 296: 1129--1132.
\input{hauert_s02}

\bibitem{hauert_s07}
Hauert C, Traulsen A, Brandt H, Nowak MA, Sigmund K (2007) Via freedom to
  coercion: The emergence of costly punishment.
\newblock Science 316: 1905--1907.
\input{hauert_s07}

\bibitem{camerer_03}
Camerer CF (2003) Behavioral Game Theory: Experiments in Strategic Interaction.
\newblock Princeton: Princeton University Press.
\input{camerer_03}

\bibitem{grujic_pone10}
Gruji{\'c} J, Fosco C, Araujo L, Cuesta JA, S{\'a}nchez A (2010) Social
  experiments in the mesoscale: Humans playing a spatial prisoner's dilemma.
\newblock PLoS ONE 5: e13749.
\input{grujic_pone10}

\bibitem{grujic_pone12}
Gruji{\'c} J, R{\"o}hl T, Semmann D, Milinksi M, Traulsen A (2012) Consistent
  strategy updating in spatial and non-spatial behavioral experiments does not
  promote cooperation in social networks.
\newblock PLoS ONE 7: e47718.
\input{grujic_pone12}

\bibitem{grujic_srep12}
Gruji{\'c} J, Eke B, Cabrales A, Cuesta JA, S{\'a}nchez A (2012) Three is a
  crowd in iterated prisoner's dilemmas: experimental evidence on reciprocal
  behavior.
\newblock Sci Rep 2: 638.
\input{grujic_srep12}

\bibitem{gracia-lazaro_pnas12}
Gracia-L{\'a}zaro C, Ferrer A, Ruiz G, Taranc{\'o}n A, Cuesta J, et~al. (2012)
  Heterogeneous networks do not promote cooperation when humans play a
  prisoner's dilemma.
\newblock Proc Natl Acad Sci USA 109: 12922--12926.
\input{gracia-lazaro_pnas12}

\bibitem{gracia-lazaro_srep12}
Gracia-L{\'a}zaro C, Cuesta J, S{\'a}nchez A, Moreno Y (2012) Human behavior in
  prisoner's dilemma experiments suppresses network reciprocity.
\newblock Sci Rep 2: 325.
\input{gracia-lazaro_srep12}

\bibitem{rand_n12}
Rand D, Greene J, Nowak M (2012) Spontaneous giving and calculated greed.
\newblock Nature 489: 427--430.
\input{rand_n12}

\bibitem{santos_md_jtb12}
Santos M, Pinheiro F, Santos F, Pacheco J (2012) Dynamics of $n$-person
  snowdrift games in structured populations.
\newblock J Theor Biol 315: 81--86.
\input{santos_md_jtb12}

\bibitem{maynard_82}
Maynard~Smith J (1982) Evolution and the Theory of Games.
\newblock Cambridge, U.K.: Cambridge University Press.
\input{maynard_82}

\bibitem{fischbacher_ee07}
Fischbacher U (2007) \protect{z-Tree: Zurich} toolbox for ready-made economic
  experiments.
\newblock Exp Econ 10: 171--178.
\input{fischbacher_ee07}

\end{thebibliography}
\end{document}